\documentclass[aps,twocolumn,amsmath,showkeys,showpacs,superscriptaddress,prb]{revtex4-2}
\usepackage{dcolumn}
\usepackage{bm}
\usepackage[T1]{fontenc}
\usepackage{amsmath}
\usepackage{amssymb}
\usepackage{graphicx}
\usepackage{tabularx}
\usepackage{orcidlink}
\usepackage{hyperref}
\hypersetup{colorlinks=true, linkcolor=blue, citecolor=blue, urlcolor=blue, pdftitle={Controlling the Anomalous Hall Conductivity by (111)-Epitaxial Strain in the Mn$_3$NiN Antiperovskite}, pdfauthor={D. Torres-Amaris}}

\usepackage[colorinlistoftodos]{todonotes}

\definecolor{emerald}{rgb}{0.31, 0.78, 0.47}

\begin{document}

\title{Anomalous Hall conductivity control in Mn$_3$NiN antiperovskite by epitaxial strain along the kagome plane}

\author{D. Torres-Amaris\,\orcidlink{0000-0003-0303-2334}}
\email{daniel.torres@saber.uis.edu.co}
\affiliation{School of Physics, Universidad Industrial de Santander, Carrera 27 Calle 09, 680002, Bucaramanga, Colombia}

\author{A. Bautista-Hernandez\,\orcidlink{}}
\affiliation{Facultad de Ingenier\'ia, Benemérita Universidad Autónoma de Puebla, Apartado Postal J-39, Puebla, Pue. 72570, México.}
 
\author{Rafael González-Hernández\,\orcidlink{0000-0003-1761-4116}}
\affiliation{Departamento de Física y Geociencias, Universidad del Norte, Km. 5 Vía Antigua Puerto Colombia, Barranquilla 080020, Colombia.}
 
\author{Aldo. H. Romero\,\orcidlink{0000-0001-5968-0571}}
\affiliation{Department of Physics and Astronomy, West Virginia University, WV-26506-6315, Morgantown, United States}
\affiliation{Department of Physics and Materials Science, University of Luxembourg, 1511 Luxembourg, Luxembourg.}
\affiliation{Facultad de Ingenier\'ia, Benemérita Universidad Autónoma de Puebla, Apartado Postal J-39, Puebla, Pue. 72570, México.}

\author{A. C. Garcia-Castro\orcidlink{0000-0003-3379-4495}}
\email{acgarcia@uis.edu.co}
\affiliation{School of Physics, Universidad Industrial de Santander, Carrera 27 Calle 09, 680002, Bucaramanga, Colombia}

\begin{abstract}  
Antiferromagnetic manganese-based nitride antiperovskites, such as Mn$_3$NiN, hold a triangular frustrated magnetic ordering over their kagome lattice formed by the Mn atoms along the (111)-plane. As such, frustration imposes a non-trivial interplay between the symmetric and asymmetric magnetic interactions, which can only reach equilibrium in a noncollinear magnetic configuration. Consequently, the associated electronic interactions and their possible tuning by external constraints, such as applied epitaxial strain, play a crucial role in defining the microscopic and macroscopic properties of such topological condensed matter systems. 
Thus, in the present work, we explored and explained the effect of the epitaxial strain imposed within the (111)-plane, in which the magnetic and crystallographic symmetry operations are kept fixed, and only the magnitude of the ionic and electronic interactions are tuned. 
We found a linear shifting in the energy of the band structure and a linear increase/decrease of the available states near the Fermi level with the applied strain. Concretely, the compression strain reduces the Mn-Mn distances in the (111) kagome plane but linearly increases the separation between the stacked kagome lattices and the available states near the Fermi level.
Despite the linear controlling of the available states across the Fermi energy, the anomalous Hall conductivity shows a non-linear behavior where the $\sigma_{111}$ conductivity nearly vanishes for tensile strain. On the other hand, $\sigma_{111}$ fetches a maximum increase of 26\% about the unstrained structure for a compression value close to $-$1.5\%.
This behavior found an explanation in the non-divergent Berry curvature within the kagome plane, which is increased for constraining but significantly reduced for expansion strain values. Our results indicate a distinct correlation between the anomalous Hall conductivity and the Berry curvature along the (111)-plane as a function of the strain. The Berry curvature acts as the source and the strain as the control mechanism of this anomalous transport phenomenon.
\end{abstract}


\maketitle

\section{Introduction}
Antiperovskites, $A_3BX$ \cite{Krivovichev2008anti} (also known as inverse-perovskites) such as Mn$_3$NiN, are cubic structures similar to perovskites in which the cation and anion position are interchanged within the unit cell for the standard perovskite crystal structure. 
Thus, the anions occupy the octahedral center instead of the corners, which becomes the site for the transition metal cations, forming a $XA_6$ octahedra.
This type of coordination, coupled with their tangible magnetic response, gives unique properties such as the anomalous Hall effect. \cite{PhysRevMaterials.3.094409,gurung2019anomalous}, negative thermal expansion \cite{peng2013mn}, giant piezomagnetism \cite{boldrin2018giant}, magnetic frustration \cite{Fruchart1978,zemen2017frustrated}, among others \cite{Wang2019,Zhao2012,Takenaka2012,Garcia-Castro2020, Garcia-Castro2019, Boyer1990,PhysRevB.78.184414,doi:10.1080/00150193.2017.1283171,Fiebig2016,spaldin2010,wu_main_2021,kim_high-efficiency_2020}. 
In particular, the magnetic frustration in triangular magnetic coordination relies on the Mn--Mn exchange and the Mn--N--Mn superexchange interaction. 
Thus, the metallic RKKY interaction, which favors an antiferromagnetic collinear arrangement \cite{10.1143/PTP.16.45}, is more prominent but is in competition with the superexchange \cite{PhysRev.100.564,KANAMORI195987} interaction mediated by the nitrogen at the cell center. 
Apart from the exchange and the superexchange, there is also an antisymmetric coupling, the Dzyaloshinskii-Moriya interaction  (DMI), which is present in the (111)-plane inducing canting, which further increases the frustration \cite{PhysRevB.83.214406,zemen2017frustrated}. 
Combining all the discussed interactions converges into non-trivial, noncollinear, and possibly chiral magnetic ordering. In this case, the chirality is of vectorial nature and comes from the removal of the mirror symmetry due to the magnetic orderings, developing a well defined handedness given by $k=\frac{2}{3\sqrt{3}}\sum_{ij}[\vec{S_i}\times\vec{S_j}]$ (where $i,j$ runs over all the magnetic moments) \cite{grohol_spin_2005,PhysRevB.78.144404}.  
For example, the triangular frustrated magnetism in Mn$_3$NiN is compatible with the $\Gamma_{5g}$ and  $\Gamma_{4g}$ \cite{Fruchart1978} magnetic orderings. 
The $\Gamma_{4g}$ ordering is symmetric under the simultaneous application of both the time-reversal symmetry, $\mathcal{T}$, and the mirror symmetry, $\mathcal{M}$, while in the $\Gamma_{5g}$ ordering, the $\mathcal{T}*\mathcal{M}$ is broken \cite{gurung2019anomalous}. 
Thus, in absence of magnetic ordering, the crystallographic symmetry of the antiperovskite belongs to the cubic \emph{Pm$\bar{3}m$} (SG. 221). Nonetheless, once the magnetic ordering is considered in the structure, the overall symmetry is reduced to the rhombohedral \emph{R$\bar3$m'} (MSG. 166.101) and \emph{R$\bar3$m} (MSG. 166.97) for the $\Gamma_{4g}$ and  $\Gamma_{5g}$, respectively. Here, the \emph{R$\bar3$m} has mirror symmetry over the \textit{M}$_{100}$, \textit{M}$_{110}$ and \textit{M}$_{010}$ planes in the Seitz notation, meanwhile, the \emph{R$\bar3$m'} needs and additional $\mathcal{T}$ after the mentioned mirror operations.
In the Mn$_3$NiN case, both magnetic orderings would present similar magnetocrystalline anisotropic energy, corroborated experimentally \cite{bertaut1972rotation,fruchartMn3SnN1977}. 
Moreover, although the overall electronic band structure is nearly identical for both orderings, the mirror symmetry breaking in the $\Gamma_{4g}$ would induce tangible properties associated with the spin polarization within the electronic states.
One prominent example of such properties is the anomalous Hall effect (AHE) in non-centrosymmetric crystals, a property accessible with the $\Gamma_{4g}$ antiferromagnetic noncollinear ordering but prohibited in the $\Gamma_{5g}$ case \cite{gurung2019anomalous}.

The family of manganese nitride antiperovskites, capable of holding any of the mentioned frustrated magnetic orderings, represents the perfect prototype candidates to investigate the anomalous Hall conductivity (AHC), its source, and possible control mechanisms. 
Moreover, the strong magnetostructural coupling present in the Mn-based antiperovskites \cite{pulkkinen2020coulomb,hobbs2003understanding}, on top of the subtle balance between the magnetic, ionic, and electronic structure, opens the door to engineering a route for AHC controlling using external constraints. 
Despite several theoretical studies that had already been dedicated to exploring the controllability of the AHC in Mn$_3$NiN by other authors \cite{gurung2019anomalous,huyen2019topology,PhysRevMaterials.3.094409,Boldrin2019,huyen2019topology}, the source of its behavior is not yet completely understood. 
For example, the reported AHC for Mn$_3$NiN ranges from $\sigma_{AHE}$= 130 S$\cdot$cm$^{-1}$ \cite{gurung2019anomalous} to $\sigma_{AHE}$ = 375.7 S$\cdot$cm$^{-1}$ \cite{huyen2019topology}, a very wide range likely related to a strong dependence on the calculations approach and parameters choosen by the authors.   
Additionally, the AHC in (001) Mn$_3$NiN strained thin films has been reported as $\sigma_{AHE}\approx$ 170 S$\cdot$cm$^{-1}$ based-on theoretical calculations whereas the measured value is around 22 S$\cdot$cm$^{-1}$ \cite{Boldrin2019}. 
Despite the disagreement, when the measured and computed AHC are compared, their conclusion was related to a direct compression (tension) to increase (decrease) relation between the strain and the AHC. 
However, applying the strain in such a direction would distort the crystal into a tetragonal symmetry, changing the magnetic and structural relationship and adding a weak-ferromagnetic canting into the magnetic response \cite{boldrin2018giant}. 
Because the symmetry is the trigger of the AHC in the antiferromagnetic antiperovskites, distortions and changes in symmetry operations could put a veil over the actual control mechanism and physical origins of this property in Mn$_3$NiN.  
Moreover, the experimental epitaxial growth of thin films of antiperovskites onto perovskites, SrTiO$_3$, has been achieved \cite{quintela2020epitaxial,quintela2017epitaxial} demonstrating the feasibility of obtaining thin-films. 
Another experimental study was performed, achieving an epitaxial growth of Mn$_3$NiN on the piezoelectric BaTiO$_3$ \cite{johnson2021strain} showing a correlation between the strain and the AHC values. 
So far, it is possible to experimentally explore the effect of the strain on antiferromagnetic nitride antiperovskites such as Mn$_3$GaN and Mn$_3$NiN. 
Despite these efforts, the physics behind the entanglement between the electronic, magnetic, and structural degrees of freedom is delicate. The explanation of the strain effect still allows room for a more profound comprehension. 
Therefore, in this paper, we show from first-principles calculations the analysis of the epitaxial strain as a control parameter of the electronic properties. Furthermore, we explored the AHC, and the Berry curvature, BC, features in the antiferromagnetic antiperovskite Mn$_3$NiN as a prototype among its family. Thus, our results explain the physical origin of the AHC controllability and possible experimental tuning.

This paper is organized as follows: We present the computational details and theoretical approaches required for the analysis of structural and electronic phenomena within the Mn$_3$NiN antiperovskite (in Section \ref{sec2}). Then, we condensed our results and research associated with the structure behavior, electronic properties, the anomalous Hall conductivity, and the Berry curvature (in Section \ref{sec3}).
Finally, we show our conclusions, in Section \ref{conclusions}, and general remarks.

\begin{figure*}[]
        \centering
        \includegraphics[width=16.0cm,keepaspectratio=true]{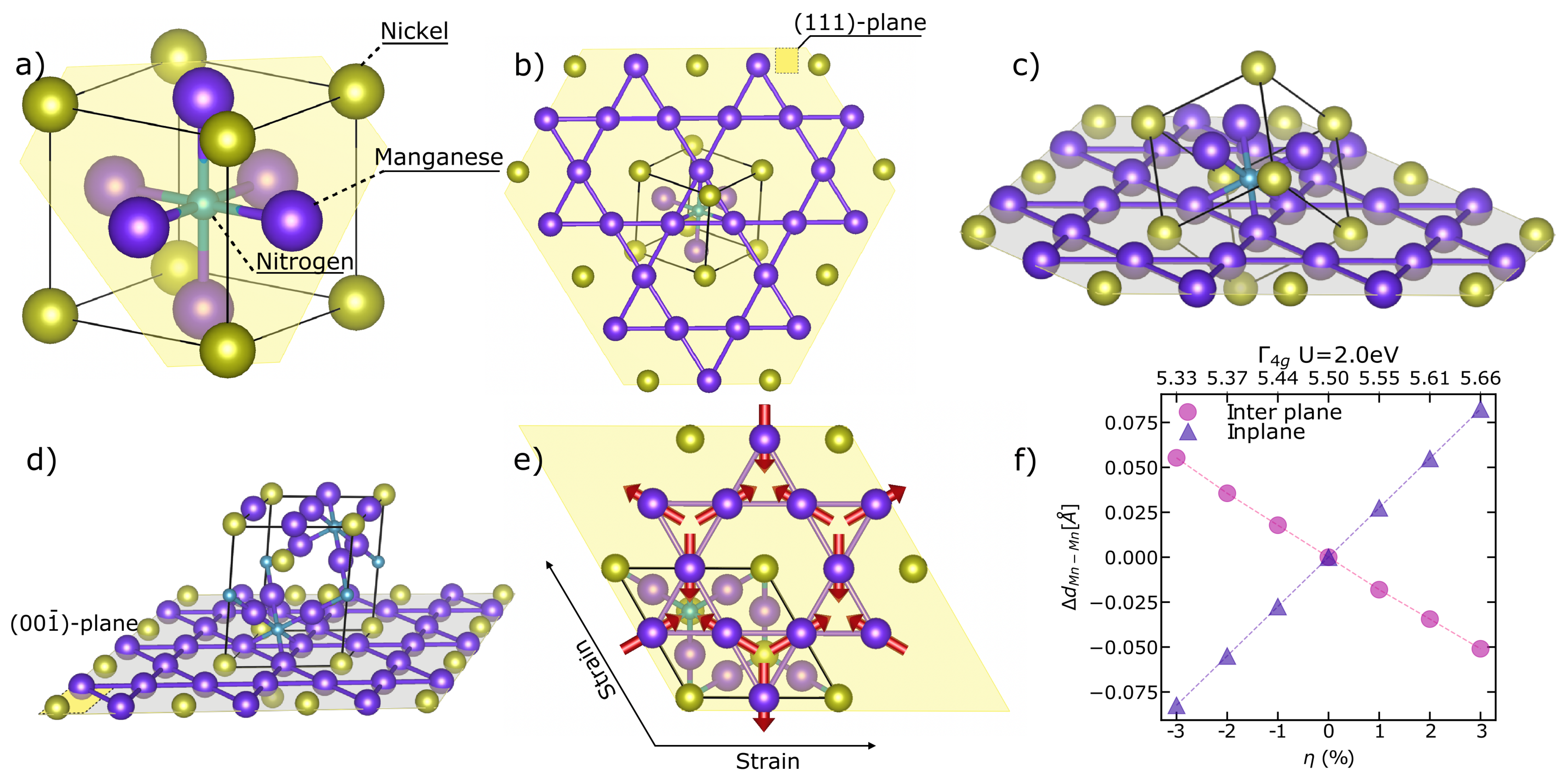}
        \caption{(Color online) (a) Mn$_3$NiN cubic structure shows the Mn, Ni, and N atoms in violet, yellow, and green colors. (b) kagome lattice on the (111)-plane formed by the Mn atoms highlighted on top of the cubic Mn$_3$NiN structure. (c)  Structure rotation aligning the (111)-plane kagome lattice with the $xy$-plane. (d) 15-atoms Hexagonal cell is recognized in the new orientation. (e) $\Gamma_{4g}$ magnetic orderings, in red, shown top of the kagome lattice and the schematics of the strain application. (f) Variation of the inter-plane and in-plane Mn--Mn distances as a function of the epitaxial strain.}
        \label{structure}
\end{figure*}

\section{Computational Details}\label{sec2}
We performed first-principles calculations within the density-functional theory (DFT) \cite{PhysRev.136.B864,PhysRev.140.A1133} approach by using the \textsc{vasp} code (version 5.4.4) \cite{Kresse1996,Kresse1999}. 
The projected-augmented waves scheme, PAW \cite{Blochl1994}, was employed to represent the valence and core electrons. 
The electronic configurations considered in the pseudo-potentials, as valence electrons, are Mn: (3$p^6$3$d^5$4$s^2$, version 02Aug2007), Ni: (3$p^6$3$d^8$4$s^2$, version 06Sep2000), and N: (2$s^2$2$p^3$, version 08Apr2002). 
The exchange-correlation was represented within the generalized gradient approximation GGA-PBEsol parametrization \cite{Perdew2008}. The Mn:3$d$ orbitals were corrected through the DFT$+U$ approximation within the Liechtenstein formalism \cite{Liechtenstein1995}.  
Due to the strong magnetostructural response observed in the Mn$_3A$N antiperovskites \cite{Singh2021}, we used the $U$ = 2.0 eV parameter in the Mn:3$d$ orbitals. This $U$ value allows the structural optimization to reproduce the experimentally observed lattice parameter, which is key in this case to obtain an appropriate charge distribution and, ultimately, the electronic properties under strain. Moreover, we also compared the PBEsol$+U$ obtained electronic structure of Mn$_3$NiN with the computed by the strongly constrained and appropriately normed semilocal density functional, SCAN, \cite{PhysRevLett.115.036402,florez2022spin-phonon}, observing a fair agreement of the electronic structure in both cases. Importantly, recent reports of SCAN-based calculations have shown results in good agreement with the experimental reports, including lattice constant \cite{PhysRevMaterials.2.095401,buda2017characterization}, the magnetic and the electronic structure \cite{pulkkinen2020coulomb} in strongly-correlated 3$d$ perovskites and Heusler Mn-based alloys \cite{barbiellini2019}.
All the procedures described above are essential due to the needed accuracy related to the lattice degrees of freedom as a function of the applied strain and its effect on the magnetostructural behavior.
The periodic solution of the crystal was represented by using Bloch states with a Monkhorst-Pack \cite{PhysRevB.13.5188} \emph{k}-point mesh of 12$\times$12$\times$12 and 600 eV energy cut-off to give forces convergence of less than 0.001 eV$\cdot$\r{A}$^{-1}$ and an error in the energy less than ~10$^{-6}$ eV.  
The spin-orbit coupling (SOC) was included to consider non-collinear magnetic configurations \cite{Hobbs2000}.  
The anomalous Hall conductivity, and associated observables, were obtained with the Python library  \textsc{WannierBerri} \cite{tsirkin2021high} using the maximally localized Wannier functions and the tight-binding Hamiltonian generated with the \textsc{Wannier90} package \cite{Pizzi2020}. 
The interpolation was performed with 80 Wannier functions with projections on the \textit{s},\textit{p},\textit{d} orbitals for the Mn and Ni atoms and \textit{s},\textit{p} for N atoms. For the disentanglement process, we used an energy window $+$3.0 eV higher than Fermi level as the maximum, and none for the minimum, and a convergence tolerance of 5.0$\times10^{-8}$ \AA$^2$. 
The atomic structure figures were elaborated with the \textsc{vesta} code \cite{vesta}.
Finally, the band structure was analyzed with the Python library \textsc{PyProcar} \cite{HERATH2020107080}.

\section{Results and Discussion:}\label{sec3}
As commented before, we aim to avoid additional contributions induced by the strain application in the (001)-plane of the 5-atom reference, depicted in Fig \ref{structure}(a). Therefore, we studied the effect of the epitaxial strain applied only in the directions parallel to the (111)-plane so that the kagome lattice, magnetic ordering, and their associated symmetry conditions are conserved.
To observe the atomic arrangement present in the (111)-plane and its spatial orientation, a broader view of that zone is shown in Fig. \ref{structure}(b). Here, the kagome lattice formed by the Mn atoms is highlighted in a yellow plane. 
To gain access to the (111)-plane of the 5-atom reference, the structure was rotated to make that plane parallel to the cartesian $xy$-plane as shown in Fig. \ref{structure}(c).
In such orientation, it is obtained an equivalent 15-atom hexagonal cell shown in Fig. \ref{structure}(d). This achieves a better representation of the structural symmetry and allows the homogenous application of the epitaxial strain.
In this new representation, the \textit{a} direction of the lattice belongs in the $xy$-plane and serves as a linearly independent crystallographic direction to apply the epitaxial strain.
As it can be seen in Fig. \ref{structure}(e), stretching along the plane would only change the Mn--Mn distance and not the atomic arrangement.
In this setup, the strain controls the intensity of the exchange and the superexchange interactions only by modifying the interatomic distances but conserving the magnetic symmetry.  

Concretely, the strain was applied as follows. The \textit{a} lattice parameter is variated along with the values from $-$3\% to $+$3\%. Still allowing the full relaxation of the crystal structure and atomic positions along the \textit{c} direction in all cases. The applied strain percentage, $\eta$, is defined in terms of the unstrained lattice parameter $a_0$ and the imposed value $a$ as:

\begin{equation}
\label{eq:strain}
\eta = \frac{a-a_0}{a_0}\times100\%,
\end{equation}

As such, the above relationship, Eq. \ref{eq:strain}, is giving compression and tension over the structure for negative and positive values of $\eta$, respectively. 

Since the electronic, magnetic, and crystalline structures of Mn$_3$NiN are strongly entangled due to its sizeable magnetostructural coupling, the cell optimization and electronic relaxation were carefully performed within the PBEsol$+U$ approximation. 
The latter in order to reproduce the experimental unstrained cell lattice constant ($a_0$ = 3.886 \r{A} below $T_N$ = 262 K) \cite{NA20113447} with a stable $\Gamma_{4g}$ magnetic ordering and to obtain a correct relaxed structure under strain. 
The best agreement between the experiment measured and the computed lattice parameter was found for $U$ = 2.0 eV (see Table \textcolor{blue}{S1} in the Supplemental Material).  This correction leads to a hexagonal cell with the parameters of \textit{a} = 5.496 \r{A} and  \textit{c} = 6.726 \r{A} within the stable $\Gamma_{4g}$ magnetic ordering. This value of lattice parameters are equivalent to a lattice parameter $a_0$ = 3.885 \r{A} in the 5-atom reference. 
Therefore, the Hubbard correction and the volume cell optimization helped to avoid a pre-strained setup which is the case of pure LDA/PBE based calculations, in which a volume adjustment is needed because of the under/overestimation of the experimentally observed value \cite{buda2017characterization}. 
Moreover, recent studies indicate a strong dependence on the electronic and lattice degrees of freedom in Mn-based compounds \cite{florez2022spin-phonon,pulkkinen2020coulomb}. 
The structural stability of the $\Gamma_{4g}$ phase of Mn$_3$NiN was tested under epitaxial strain by obtaining the full phonon-dispersion curves at $\eta$ = $-3$, $0$, and $+3$\%, see Fig. \textcolor{blue}{S1}(a). 
The latter aims to ensure that the ranges of strain chosen were within the limits of the structural stability and that no phase transitions might be induced.
As shown in Fig. \textcolor{blue}{S1}(a), the full phonon-dispersion shows no imaginary or unstable phonons, confirming the structural and vibrational stability of Mn$_3$NiN under the considered strain values.
Furthermore, the magnetic phase stability, of the $\Gamma_{4g}$ over the $\Gamma_{5g}$, was also tested (see Fig. \textcolor{blue}{S1}(b)). As it can be observed from Fig. \textcolor{blue}{S1}(b), the compression epitaxial strain reinforces and stabilizes the $\Gamma_{4g}$ order, whereas, for expansion strain values, the difference in energy between the antiferromagnetic orderings is reduced. 
Thus, the negative strain values serve as a mechanism to freeze the $\Gamma_{4g}$ ordering in the Mn$_3$NiN.

\begin{figure}[t]
 \centering
 \includegraphics[width=8.7cm,keepaspectratio=true]{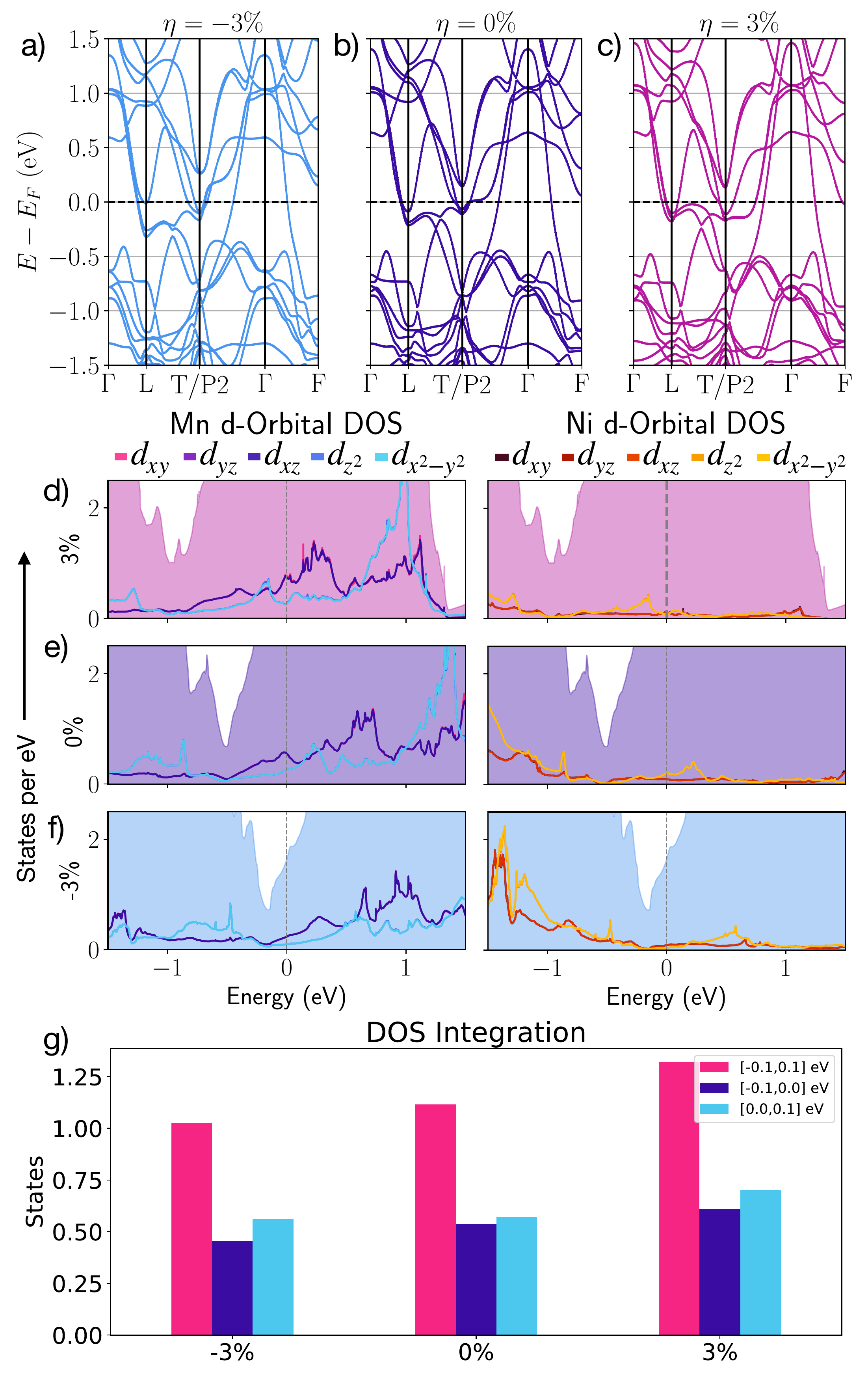}
 \caption{(Color online) (a-c) band structure of Mn$_3$NiN $\Gamma_{4g}$ non-collinear magnetic phase. (d-f) Each set showing $-3\%$, $0\%$ and $+3\%$ strain cases DOS and d-orbital projected DOS. (g) DOS integration 0.1 eV under, over and around Fermi energy.}
 \label{electronic}
\end{figure} 

The variations of the plane-to-plane distance between the kagome planes, in comparison with the distance between two Mn nearest atoms of the same plane, are presented in Fig. \ref{structure}(f).
The compression/tension strain application, directly on the plane of the magnetic kagome lattice, produces the expected response of increasing/decreasing the separation of the (111) family of planes, as shown in the graph of Fig. \ref{structure}(f). 
Furthermore, linear variations of the distance between Mn atoms of the same plane induced inverse linear variation in separating the kagome planes. 
Consequently, the exchange and superexchange interactions can be finely tuned, ultimately gaining control over the frustration mechanism.
Thus, the (111) applied strain is advantageous from the symmetry point of view because the initial \emph{R$\bar3$m} symmetry is preserved along the deformation path on the 15-atom reference. 
Therefore, the symmetry relationships do not change; instead, only the electronic effects can be tuned through the control over the interacting moments by the epitaxial strain. 
Additionally, symmetry preservation allows straining and optimizing on the 15-atom reference and then returning to the 5-atom representation to perform the rest of the calculations and analysis, avoiding electronic bands unfolding issues. 
To recover the 5-atom representation, we made use of the transformation matrices, as implemented in the \textsc{findsym} tool \cite{stokes2005findsym,stokes2017findsym}. In what follows, all the calculations and analyses are carried out on the 5-atom reference for each relaxed strain cell.

The electronic band structure calculated along the $\Gamma$--$L$--$T$/$P2$--$\Gamma$--$F$ path in the BZ for the $\eta$ = $-$3\%, 0\%, and $+$3\% in the 5-atom reference is presented in Fig. \ref{electronic}(a-c). Here, the $L$--$T$ path lies along the (111)-plane (where the kagome lattice lies) while the $P2$--$\Gamma$ path runs perpendicularly to the same plane. When the structure is compressed, the energy bands close to the Fermi energy are pushed down. This is more noticeable along the $L$--$T$ branch, as shown in Fig.~\ref{electronic}(a). Meanwhile, as shown in the same figure, electron bands shift up in energy together in the $P2$--$\Gamma$ path. On the other hand, the behavior is the opposite when tension is applied, as observed in the $\eta$ = +3\% case presented in Fig. \ref{electronic}(c).

Projections per atomic specie of the electronic band structure for $\eta$ = $-$3\%, 0\%, and 3\% (see Fig. \textcolor{blue}{S2} in the Supplemental Material) show that Mn:$3d$-states represent the major contribution above the Fermi level. 
Meanwhile, Ni:$3d$-states dominate the band structure under the Fermi energy, with its most substantial contribution around -1.25 eV. 
Both Mn and Ni atomic species share the intermediate ($-$0.5, $+$0.5) eV range of energy. Thus, the conduction phenomena result from the hybridization of the Mn and Ni $d$-orbitals around the Fermi energy.
Aiming to analyze the available charge and states around the Fermi level, we computed the DOS for $\eta$ = $-$3\%, 0\%, and $+$3\% and the results are contained in Fig. \ref{electronic}(d-f).  
Here, we observed a linear shift with respect to the energy of the total DOS as a response to the applied strain. Tracking the minimum of the DOS, located at $-$0.5 eV in the $\eta$ = 0\% DOS plot in Fig. \ref{electronic}(e), which moves up (down) in energy for compression (tension), this behavior becomes clear. More precisely, the available states near the Fermi level decrease with compression and increase with tension, see Fig. \ref{electronic}(g).
To further dive into the DOS subtleties around the Fermi level, the Mn and Ni:3$d$-orbitals projections of the DOS are included in Fig. \ref{electronic}(d-f).
As it can be observed, the contribution at the Fermi level from the 3$d_{xy/yz/xz}$ orbitals increase (decrease) for tension (compression) strain values. The same is the case for 3$d_{z^2/x^2-y^2}$.
In general, the 3$d$ orbitals are pushed upwards in energy when the structure is compressed.
In the case of Ni:3$d$ orbitals, a marginal contribution is observed close to the Fermi level.
Finally, direct integration of the total DOS for each $\eta$ in the [$-$0.1,$+$0.1] eV interval, as presented in Fig. \ref{electronic}(g), confirms the relationship between the electronic states and strain inferred from the complete and partial DOS analysis. 
Moreover, the integration of the DOS over the ranges [$-$0.1,0.0] and [0.0,$+$0.1] eV for the occupied and unoccupied bands, respectively, follow the same behavior already observed in the [$-$0.1,$+$0.1] eV interval. 

Before discussing our AHC findings, it is worth mentioning the different sources behind the AHC. 
The AHC in crystals can be the result of different sources: the intrinsic, side jump, and the skew scattering contribution, as shown in Ref. \cite{nagaosa2010anomalous}. 
The last two of them are a consequence of impurities in the crystal that deflect and scatter the electrons sideways. 
The intrinsic contribution of the AHC arises from the interband coherence induced by an external magnetic field. 
In this work, we concentrate our attention on the intrinsic component of the AHC, which results from the electronic, magnetic, and structural properties of a perfect crystal. 
Additionally, the intricate combination of the many interactions present in the frustrated triangular shape created between the Mn atoms in the kagome place reduces the symmetry to \emph{R$\bar3$m} in the $\Gamma_{5g}$ case. Finally, in the case of the $\Gamma_{4g}$ magnetic ordering, shown in Fig \ref{structure}(e), the $\mathcal{M}$-symmetry is also removed, ending up with the \emph{R$\bar3$m'} symmetry. 
This lack of $\mathcal{M}$-symmetry is essential for the existence of AHC in the $\Gamma_{4g}$ phase. 
The $\mathcal{M}$-symmetry is also the reason for the absence of that property in the $\Gamma_{5g}$ phase.
Thus, the AHC reported in this work is calculated based on the relationship defined as follows \cite{nagaosa2010anomalous}:

\begin{equation}
\label{eq:ahc}
    \sigma^{AHC}_{\alpha\beta}=-\frac{e^2}{\hbar}\epsilon_{\alpha\beta\gamma} \int_{BZ}\sum_n\frac{d^3\vec{k}}{(2\pi)^3} f_n(\vec{k})\Omega_n ^\gamma(\vec{k}),
\end{equation}
 
\begin{figure}[t]
 \centering
 \includegraphics[width=9.0cm,keepaspectratio=true]{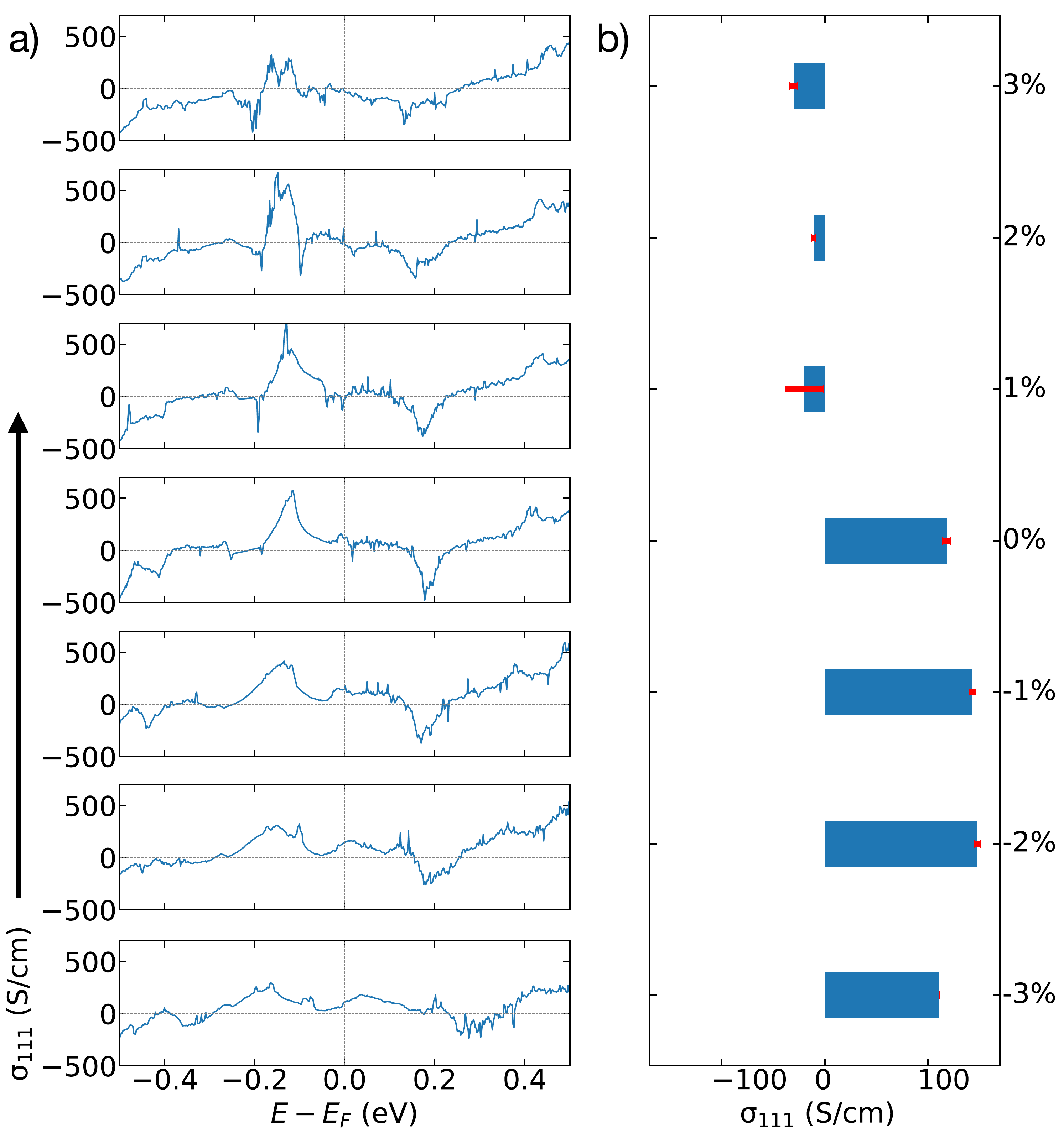}
 \caption{(Color online) (a) Anomalous Hall conductivity as a function the Fermi energy for various strain values, showing a downward shifting behaviour marked with a red dashed line. (b) AHC value at Fermi energy with the the error as the standard deviation of the last 20 iterations included as a red bar for each strain value}
 \label{ahc}
\end{figure} 
 
The latter, Eq. \ref{eq:ahc}, as implemented in the \textsc{WannierBerri} code \cite{tsirkin2021high}. 
Here, in Eq. \ref{eq:ahc}, $\epsilon_{\alpha\beta\gamma}$ is the antisymmetric tensor, $\sum_n f_n(\vec{k})\Omega_n ^\gamma(\vec{k})$ is the summation over all the included bands contribution to the Berri curvature, $\Omega_{\gamma}(\vec{k})$, and $f_n(\vec{k})$ is the Fermi distribution. 
In Eq. \ref{eq:ahc} the $\gamma$ subscript runs over a discrete grid of energy points, allowing the AHC calculation in other energy levels apart from the Fermi level. 
By looking at the Eq. \ref{eq:ahc}, two main factors are candidates to explain the AHC behavior as a function of the epitaxial strain: The available electronic states around the Fermi level and the Berry curvature integration in the BZ.
The $\Gamma_{4g}$ phase of the Mn$_3$NiN system is a non-collinear antiferromagnet with a non-zero magnetic moment of each Mn atom but with zero net magnetization. Consequently, without a net internal or external magnetic field, the Hall conductivity must result from the anomalous Hall effect (AHE) through a non-vanishing Berry curvature, $\sigma_{\alpha\beta}$ as in Eq. \ref{eq:ahc}. 
The latter is resulting in the following tensor for the \emph{R$\bar3$m'} magnetic symmetry group \cite{PhysRevB.92.155138,gurung2019anomalous}:

\begin{equation}
\label{ahc:tensor}
\sigma_{\Gamma_{4g}}=
\begin{pmatrix}
0 & \sigma_{xy} & -\sigma_{xy}\\
-\sigma_{xy} & 0 & \sigma_{xy} \\
\sigma_{xy} & -\sigma_{xy}& 0
\end{pmatrix}
\end{equation}
with all the non-zero components identical $\sigma_{xy}=\sigma_{zx}=\sigma_{yz}$ and therefore represented all by the $\sigma_{xy}$ component. 
The strain application proposed in this work is now advantageous because the symmetry preservation guarantees a fixed AHC tensor form and symmetry conditions, as seen in Eq. \ref{ahc:tensor}. 
As such, the setup for the strain, as seen in Fig. \ref{structure}(e), is the key to studying the AHC in Mn$_3$NiN as a pure function of the strain without altering the allowed symmetry features, and then, the variations on the AHC in the (111)-plane $\sigma^{\eta}_{111}=\frac{1}{\sqrt{3}}(\sigma_{xy}^{\eta}+\sigma_{yz}^{\eta}+\sigma_{zx}^{\eta})$ can be extracted as a function of $\eta$ in the kagome lattices.

Fig. \ref{ahc}(a) shows the AHC as a function of the energy, in the energy range [$-$0.5,$+$0.5] eV and for $\eta$ = $-3$, $-2$, $-1$, $0$, $1$, $2$, and $3\%$. 
In Fig. \ref{ahc}(b), is presented a barplot condensing the AHC at the Fermi level for each strain value, as well as their error bards. 
The latter, marked in red in Fig. \ref{ahc}(a), were estimated as the standard deviation of the last 20 iterations while computing the $\sigma_{xy}$ component based on the Eq. \ref{eq:ahc}. Interestingly, the AHC results show a particular behavior; its value does not just increase or decrease with the epitaxial strain; as seen in Fig. \ref{ahc}(a), the whole function suffers a flattening with the compressive and tensile stress incremental. 
Additionally, the maxima and minima of the conductivity function diverge away from the Fermi level with both types of deformation. 
Furthermore, as seen in the barplot of Fig. \ref{ahc}(b), the tensile strain produces almost an AHC vanishing value, stretching the (111)-plane as low as $\eta$ = $+$1\% and upwards reduces the conductivity dramatically. 
The compression, on the other hand, induces an increase of the AHC from $\sigma^{0\%}_{111}$ = 114 S$\cdot$cm$^{-1}$ to  $\sigma^{-1\%}_{111}$ = 144 S$\cdot$cm$^{-1}$ representing an increase of 26\%. However, the AHC remains constant in a plateau zone that holds until $\eta$ = $-$2\%. 
Further values of strain compression, after $\eta$ = $-$2\%, does not enhances the AHC, instead, the conductivity drops after this strain value, reaching $\sigma^{-3\%}_{111}$ = 111 S$\cdot$cm$^{-1}$ for $\eta$ = $-$3\%, a value similar to the case $\eta$ = 0\%. 
Looking at Fig. \ref{ahc}(a), a small spike of AHC is spotted just under the Fermi level for $\eta$ = 0\%. 
Later, it disappears for tension but enhances under compression, growing non-stop along the interval 0\%$\geq\eta\geq$$-$2\%. 
Moreover, the mentioned spike moves up to energies higher than the Fermi level, being its maximum synchronized with the Fermi level for a compressive $\eta$ between $-$1 \% and $-$2\%  
Thus, our findings suggest that the area under the AHC curve is redistributed with the strain rather than shifted. 
Considering the Eq. \ref{eq:ahc} and aiming for gathering more information on the origin of the AHC control mechanism, the $\sigma^{-1.5\%}_{111}$ and the Berry curvature were calculated. 
The conductivity for the additional strain value turned out as $\sigma^{-1.5\%}_{111}$ = 141 S$\cdot$cm$^{-1}$, confirming the plateau zone previously mentioned. 
Here, some saturation is occurring that is stable within $-$1\%$\geq\eta\geq$$-$2\%.

\begin{figure}[b]
 \centering
 \includegraphics[width=8.5cm,keepaspectratio=true]{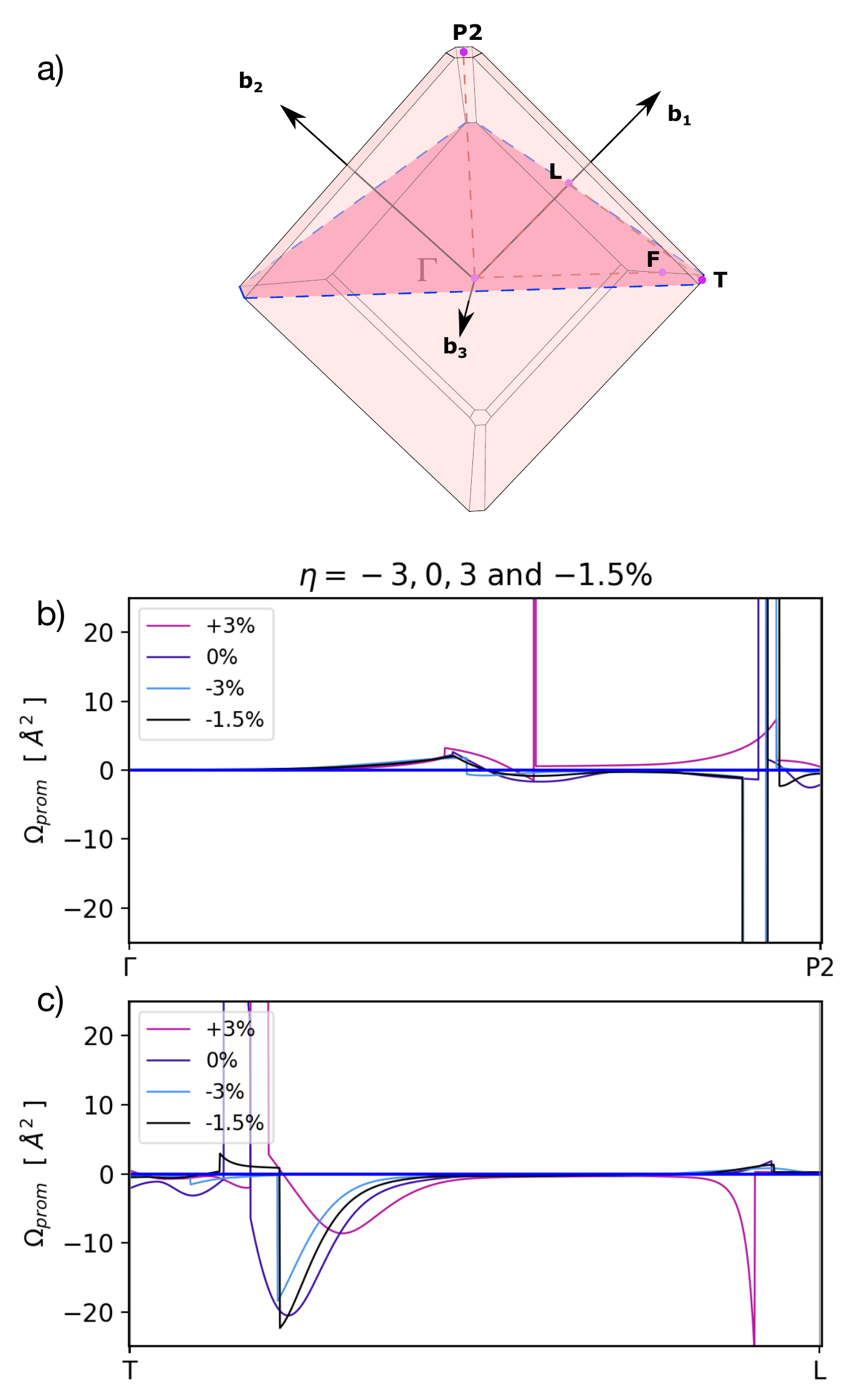}
 \caption{(Color online) Brillouine zone for the rhombohedral structure of Mn$_3$NiN is shown in part (a), highlighting a plane parallel to the (111)-plane and the high symmetry points included in the band structure calculation. The Berry curvature calculated for Mn$_3$NiN for the path connecting the high symmetry points (b) $T$--$L$ and (c) $\Gamma$--$P2$, perpendicular and within the kagome lattices, respectively.}
 \label{bc}
\end{figure}

Surprisingly, a comparison between the AHC (in Fig. \ref{ahc}) and the states available near the Fermi level (see Fig. \ref{electronic}(g)) within the range [$-$0.1,$+$0.1] eV, directly associated with $f_n(\vec{k})$, in Eq. \ref{eq:ahc}, shows no correlation. Here, we expected to find a connection because of the $\Omega_{\gamma}(\vec{k})$ dependence on the Fermi distribution.
However, the number of states increases with the tension while the AHC gets almost destroyed under such circumstances. 
A  DOS projection onto the Mn and Ni 3$d$-orbitals, which dominates most of the band structure around the Fermi level, showed a non-similar behavior to the AHC. 
The contribution of those orbitals follows the same rules as the total DOS, as already discussed in the electronic structure analysis. 
For instance, the only source of control that remains for the AHC is the Berry curvature, which will be analyzed in what follows.

The Mn$_3$NiN BZ is shown in Fig. \ref{bc}(a) in which $\Gamma$--$P2$ and $T$--$L$ are shown with respect to the (111)-plane kagome lattice.
The BC integration results for $\eta$ = $-$3.0\%,$-$1.5\%, 0.0\% and 3.0\% along the $\Gamma$--$P2$ and the $T$--$L$ paths are shown in Fig. \ref{bc}(b) and \ref{bc}(c), respectively. 
Analyzing the BC along the $\Gamma$--$P2$ path shown in Fig. \ref{bc}(b), it can be identified various discontinuities belonging to Weyl points near and at the Fermi level. 
In this case, the rhombohedral symmetry preservation throughout the strain application process is advantageous because it allows the shape of the band structure to remain mostly unaltered. As a result, the number of Weyl crossings is kept constant, and they can only move up and down in energy. 
Notably, the Weyl nodes near the Fermi level produce a divergent component of the BC that scales linearly with the strain.
Despite that, the results showed an AHC that does not follow the same linearity but instead turned out to be a nonlinear function of the strain. 
Therefore, in agreement with Huyen \emph{et al.} \cite{huyen2019topology}, our results suggest that the highly localized and divergent Berry curvature, induced by the Weyl points near the Fermi level, is not the AHC primary origin.
On the other hand, the BC in the $T$--$L$ section, shown in Fig. \ref{bc}(c), provides both types of BC, localized and not localized, over the path. 
The localized BC is once more uncorrelated to the AHC data. This confirms what has already been discussed in the BC analysis along the $\Gamma$--$P2$ segment.
Interestingly, the delocalized BC correlates to the AHC for each strain value. 
The highest values of the BC are $\Omega^{0\%}_{111}$ = $-$20.5 \r{A}$^2$, $\Omega^{-1.5\%}_{111}$ = $-$22.5 \r{A}$^2$ and $\Omega^{-3\%}_{111}$ = $-$18.4 \r{A}$^2$,  while for $+$3\% a relatively small value of BC is spotted $\Omega^{+3\%}_{111}$ = $-$8.6  \r{A}$^2$. 
The AHC values in each of the mentioned cases are $\sigma^{0\%}_{111}$ = 114 S$\cdot$cm$^{-1}$, $\sigma^{-1.5\%}_{111}$ = 140 S$\cdot$cm$^{-1}$, $\sigma^{-3\%}_{111}$ = 111 S$\cdot$cm$^{-1}$ and $\sigma^{+3\%}_{111}$ = $-$27 S$\cdot$cm$^{-1}$.  
It is important to remark that the $L$--$T$ path lies in a plane parallel to the (111)-plane, and the $P2$--$\Gamma$ is parallel to the magnetic symmetry axis ($i.e.$ along the (111)-axis and perpendicular to the kagome lattice, see Fig. \ref{bc}(a)). 
Thus, as expected, the AHE occurs only over the (111)-plane (\textit{i.e.} into the kagome lattice) and not in the perpendicular direction. 
Therefore, the AHC, induced by a nonvanishing BC in the (111)-plane, conducts the carriers over the same plane where the $\mathcal{T}*\mathcal{M}$ preserving magnetic orderings are placed.
This non-divergent BC can be attributed to interband coherence induced by the electronic field \cite{nagaosa2010anomalous}. Avoided band crossings at the Fermi energy level are beneficial to the AHC due to the strong interaction of the occupied and the unoccupied bands \cite{wang2006ahc,wang2007ahc,PhysRevLett-yao2004ahc}. The latter, observed for the computed BC within the (111)-plane included in Fig. \textcolor{blue}{S3}.

\section{Conclusions and general remarks}\label{conclusions}
Through first-principles calculations and theoretical analysis, we have investigated the strain-driven controlling of AHC in Mn$_3$NiN antiperovskite. 
We found that the strain application in the (111)-plane preserves the symmetries of the system, and its band structure with them, allowing only a rigid shifting of the bands in energy. 
Such preservation keeps intact the source of the AHC, the $\mathcal{T}\mathcal{M}$ in the $\Gamma_{4g}$ magnetic ordering, leaving the AHC tensor form unchanged in each case. 
Therefore, the AHC is a function of the distance between the Mn atoms, both of the same and different kagome lattice planes. 
Our results indicate a nonlinear compression/tension and enhancing/decreasing relation between the AHC and the strain. 
Moreover, the strain induced a redistribution of the AHC function maxima and minima concerning the Fermi energy. The magnitude of the AHC and the BC as strain functions showed a correlation over their components in the kagome lattice plane. However, there is a limit to this control mechanism. 
The maximum AHC value is reached within $-$1\%$\leq\eta\leq$ $-$2\%, where further compression only reduces the AHC. 
Remarkably, neither the total nor the 3$d$-orbital projected DOS in the vicinity of the Fermi energy presented correlations to the AHC. Instead, the physics behind the tuning of the AHC relies on the non-divergent Berry curvature within the (111) kagome plane. The BC in the L-T path in this plane increases as the strain reduces the Mn-Mn distance. 
Thus, the strain in the (111)-plane proved to be an effective tool to tune the AHC in the $\Gamma_{4g}$ magnetic phase of Mn$_3$NiN.

\section*{Acknowledgements}
We thank L. Flores-Gomez for helpful comments and discussions. 
Calculations presented in this paper were carried out using the GridUIS-2 experimental testbed, being developed under the Universidad Industrial de Santander (SC3-UIS) High Performance and Scientific Computing Centre, development action with support from UIS Vicerrectoría de Investigación y Extension (VIE-UIS) and several UIS research groups as well as other funding resources.
The authors acknowledge the Texas Advanced Computing Center (TACC) at The University of Texas at Austin for providing (HPC, visualization, database, or grid).
Additionally, we acknowledge the XSEDE facilities' support, a project from the National Science Foundation under grant number ACI-1053575.
The authors also acknowledge the Texas Advanced Computer Center (with the Stampede2 and Bridges-2 supercomputers).
We also acknowledge the use of the SuperComputing System (Thorny Flat) at WVU, which is funded in part by the National Science Foundation (NSF) Major Research Instrumentation Program (MRI) Award $\#$1726534.
A.\,C.\,G.\,C. acknowledges the grant No. 2677: "Quiralidad y Ordenamiento Magnético en Sistemas Cristalinos: Estudio Teórico desde Primeros Principios" supported by the VIE – UIS.
A.B.H acknowledges the computational support extended to us by Laboratorio de Supercomputo del Sureste (LNS), Benemérita Universidad Autónoma de Puebla, BUAP, for performing heavy theoretical calculations.
The work by A.H.R. was supported by the grant DE-SC0021375 funded by the U.S. Department of Energy, Office of Science.

\bibliography{library}

\end{document}